\documentclass[10pt,conference]{IEEEtran}
\IEEEoverridecommandlockouts

\usepackage{siunitx}
\usepackage{cite}
\usepackage{amsmath,amssymb,amsfonts, amsthm}
\usepackage{tikz}
\usetikzlibrary{automata, positioning, arrows,calc}
\usetikzlibrary{quantikz}
\usepackage{algorithmic}
\usepackage{graphicx}
\usepackage{textcomp}
\usepackage{xcolor}
\usepackage{caption} 
\usepackage{subcaption}

\usepackage{hyperref}
\usepackage{mathtools}
\usepackage{multirow}
\usepackage{graphicx}
\usepackage{multicol}
\usepackage{array}
\usepackage{enumitem}

\newtheorem{cor*}{Corollary}

\newtheorem*{remark}{Remark}

\newcommand{\mpfont}{\scriptsize}

\ifx\noeditingmarks\undefined
    \newcommand{\MPworker}[2]{{\color{#1}\vrule\vrule}{\marginpar{\color{#1}\mpfont #2}}}
\else
    \newcommand{\MPworker}[2]{}
\fi

\begin{document}

\title{
Simulation of Quantum Transduction Strategies \\
for Quantum Networks

}

\makeatletter
\newcommand{\linebreakand}{%
  \end{@IEEEauthorhalign}
  \hfill\mbox{}\par
  \mbox{}\hfill\begin{@IEEEauthorhalign}
}
\makeatother

\author{

\IEEEauthorblockN{Laura d'Avossa$^{*}$, Caitao Zhan$^{\dagger}$, Joaquin Chung$^{\dagger}$, \\
Rajkumar Kettimuthu$^{\dagger}$, Angela Sara Cacciapuoti$^{*}$, Marcello Caleffi$^{*}$}

\IEEEauthorblockN{$^{*}$University of Naples Federico II (FLY: Future Communications Laboratory), Italy. \\
$^{\dagger}$Argonne National Lab, USA.}

\thanks{
This work has been funded by the European Union under Horizon Europe ERC-CoG grant QNattyNet, n.101169850. Views and opinions expressed are however those of the author(s) only and do not necessarily reflect those of the European Union or the European Research Council Executive Agency. Neither the European Union nor the granting authority can be held responsible for them.
The work has been also partially supported by PNRR MUR RESTART-PE00000001. This material is also based upon work supported by the U.S. Department of Energy, Office Science, Advanced Scientific Computing Research (ASCR) program under contract number DE-AC02-06CH11357 as part of the InterQnet quantum networking project.}

}

\maketitle

\begin{abstract}

The Quantum Internet would likely be composed of diverse qubit technologies that interact through a heterogeneous quantum network. Thus, quantum transduction has been identified as a key enabler of the Quantum Internet.
To better study heterogeneous quantum networks, the integration of a quantum transducer component into quantum network simulators has become crucial. In this paper, we extend SeQUeNCe, an open-source, discrete-event simulator of quantum networks, with a quantum transduction component along with auxiliary hardware device models and protocols. Moreover, we explore two strategies for transmitting quantum information between superconducting nodes via optical channels, with a focus on the impact of quantum transduction on the transmission process. The performance of these strategies is analyzed and compared through simulations conducted using SeQUeNCe. Our results align with theoretical predictions, offering simulation-based validation of the strategies and providing a path to accurate, larger-scale simulations of heterogeneous quantum networks. 
\end{abstract}


\section{Introduction}
\label{sec:1}

The numerous challenges in the realization of the Quantum Internet have led the scientific community to converge toward the realization of a heterogeneous network that leverages different technologies with complementary features \cite{CacCalTaf-20, DavCacCal-24-1, DavCacCal-24, DavCalWan-23}. 
Indeed, qubits can be implemented using different hardware platforms, each of them exhibiting advantages and disadvantages. 
\textit{Superconducting technology} is regarded as a very promising quantum computing platform, because superconducting qubits can be easily fabricated and their gate implementation operates at a fast speed. However, 
communication between superconducting qubits is enabled by microwave photons at cryogenic temperatures, which impedes the realization of large-scale quantum networks of this technology\cite{Wen-17}.
Conversely, \textit{photonic technology} is recognized as the most suitable technology to 
realize quantum communications at room temperature.
Indeed, optical photons weakly interact with the environment which results in low decoherence and the possibility of preserving the quantum state in long-distance transmission without relying on cryogenic temperatures\cite{CacCalTaf-20}.
Therefore, a realization of the Quantum Internet may consist of superconducting quantum nodes and optical quantum channels.
However, superconducting qubits that communicate via microwave photons, cannot directly interact with optical photons due to the enormous gap among the frequency domains.
Thus, it is mandatory to use a \textit{quantum transducer}, which is a quantum interface that converts microwave photons into optical photons and vice-versa, 
effectively enabling the interaction between superconducting qubit platforms via optical technologies\cite{CalCacBia-18, LauSinBar-20}, as schematically represented in Fig.~\ref{fig:01}. 

\begin{figure}[t!]
    \centering
   
    \includegraphics[width=0.94\linewidth]{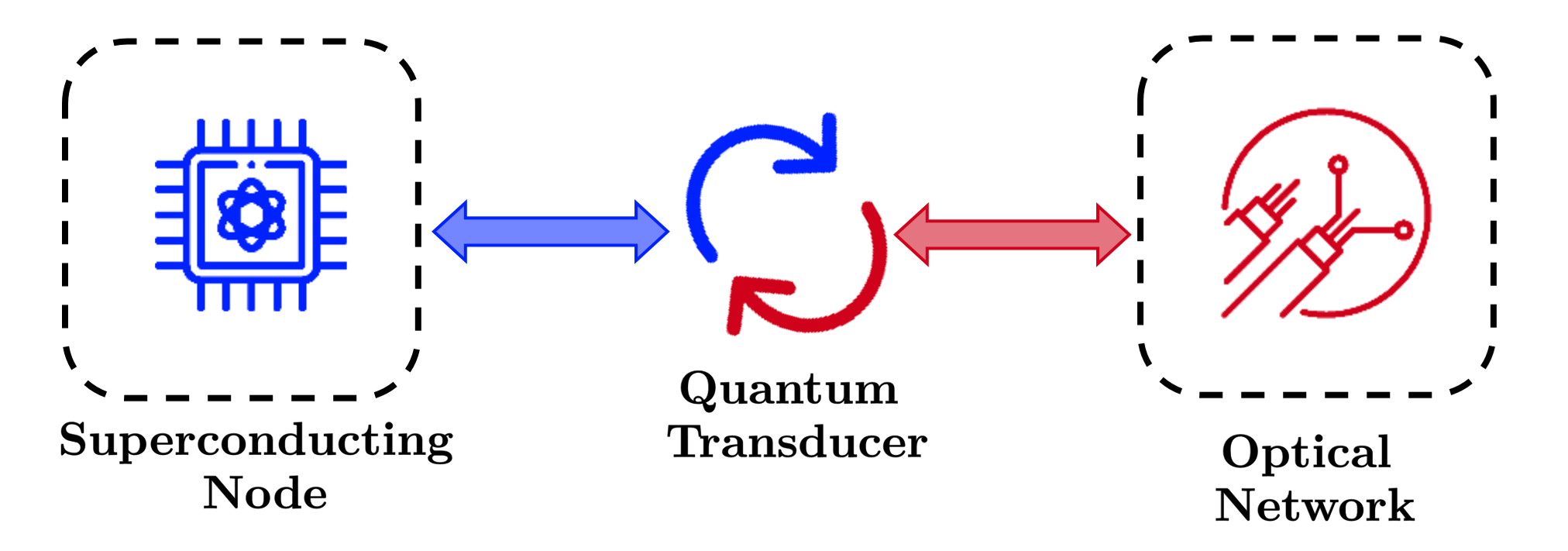}
    \caption{Quantum Transducer as an interface between superconducting nodes across an optical network.}
    \hrulefill
    \label{fig:01}
\end{figure}

Despite the great promise of superconducting qubit platforms to realize the Quantum Internet, quantum transducer hardware development is still at an early stage and 
their current efficiencies are very low.
Meanwhile, quantum network simulators~\cite{WuKolChu-21, netsquid_2021} have played a crucial role in studying quantum network hardware~\cite{ZanKolChu22,SooBenHaj24}, protocols~\cite{Dahlberg_2019,GhaZhaGup-22}, and applications~\cite{CalAmFer-24, BelKir-24}.
Thus, it is meaningful to leverage quantum network simulators to study the expected performance of quantum transducers in large-scale, heterogeneous quantum networks. 
Motivated by the above, this paper makes the following contributions:
\begin{itemize}
    \item We design and implement a new quantum transducer module in a discrete-event quantum network simulator named SeQUeNCe~\cite{WuKolChu-21}, making SeQUeNCe the first quantum network simulator to have a quantum transducer hardware module.
    \item We implement two communication strategies for quantum transducers in SeQUeNCe that enable point-to-point quantum information transfer between superconducting nodes via an optical quantum channel.
    \item Given the quantum transducer module and two strategies in SeQUeNCe, we evaluate and compare their performance via extensive simulation studies. The studies focus on the impact of quantum transduction on quantum information transmission. We open source our implementation in GitHub~\cite{Online}.
\end{itemize}

The paper is organized as follows. In Sec.~\ref{sec:02} we introduce our quantum transduction strategies that enable the quantum information transmission between remote nodes in a quantum network. 
In Sec.~\ref{sec:03} we present the design and implementation of the quantum transducer module in SeQUeNCe and then use it to implement our proposed strategies for quantum information transmission. 
In Sec.~\ref{sec:04} we show the results of our simulations,
while in Sec.~\ref{sec:05} we discuss the communication performances of the proposed strategies based on the theoretical analysis and simulation results. Finally, in Sec.~\ref{sec:06} we conclude the paper offering a glimpse into the future direction of this research topic.

\section{ 
Direct vs. Entanglement-based Quantum Transduction} 
\label{sec:02}

In a heterogeneous quantum network of superconducting nodes and optical channels, the transmission of quantum information via quantum transduction can be implemented by the following strategies: Direct Quantum Transduction (DQT) and Entanglement-based Quantum Transduction (EQT) \cite{HanFuZou-21, ZhoChaHan-22}.
In DQT, as suggested by the name, a direct transmission of quantum information is performed and the qubits are converted from one frequency to another. Conversely, the EQT strategy exploits quantum transduction for hybrid Einstein–Podolsky–Rosen (EPR) pair generation (i.e., generation of entanglement between microwave and optical photons).
Once the entanglement is successfully generated and distributed, the quantum teleportation protocol is performed.
In a nutshell, a quantum transducer requires an input laser pump to enable the frequency conversion in DQT or the entanglement generation in EQT \cite{HanFuZou-21, ZhoChaHan-22}. Different approaches can be exploited for the physical implementation of a quantum transducer. Typically, to enable the interaction of microwave and optical modes a quantum transducer consists of an optical and a microwave cavity (such as in electro-optical transducers). However, other transducer hardware platforms can exploit mediator modes --besides the microwave and optical ones-- such as mechanical or acoustic modes.
Although the choice of transducer hardware platform depends on the specific application and requirements \cite{LauSinBar-20}, the strategies for transmitting quantum information that we present here are independent of the particular hardware implementation of the transducer. 
Indeed, the most important transducer features are embedded in a single parameter, namely its \textit{conversion efficiency}, which allows us to characterize the transducer's performances, as we discuss in the next section.

\subsection{DQT: Strategy Description}
\label{sec:DQT}

For a quantum information transfer in the DQT strategy, two frequency conversions are required: (1) \textit{up-conversion} at the source node that converts a qubit from microwave to optical frequencies, and (2) \textit{down-conversion} at the destination node that converts the optical qubit back to microwave, as schematically represented in Fig.~\ref{fig:02.1} and Fig.~\ref{fig:02.2}.
However, none of these
processes are deterministic (i.e., there exists a non-zero probability that either or both conversion processes will fail) \cite{WuCuiFan-21,Tsa-11}.
The quantum transducer exhibits successful conversion probability, 
also defined as \textit{conversion efficiency} $\eta$ that strictly depends on the characteristic of the transducer hardware. 
Specifically, we denote by $\eta_\uparrow$ the up-conversion efficiency and by $\eta_\downarrow$ the down-conversion efficiency.
Despite big efforts in the realization of quantum transducers, the conversion efficiency of current hardware platforms remains well below 100\%.
Recently, bulk optomechanical transducers reached the highest photon conversion efficiency reported so far (about 50\%) \cite{AndPetPur-14, HigBurUrm-18}. 
Conversely, electro-optical transducers present lower added noise in the conversion due to the absence of mediator modes, however currently they can reach conversion efficiency values of only $10\%$ \cite{SahHeaRue-22}. As a result, obtaining high-efficiency values remains an open and crucial challenge \cite{ZhoChaHan-22}.
%in the order of $10^{-2}$ %\cite{LamRueSed-20, FanZouCha-18}. 
In the DQT protocol, the limitations given by the nondeterministic conversions are compounded by the losses of the quantum channel through which the information is transmitted.
Indeed, if quantum information successfully up-converted at source is lost or damaged by channel noise, it cannot be recovered with a copy made earlier due to the no-cloning theorem \cite{CacCalVan-20}.
For these reasons, implementing DQT is still a hard challenge for quantum information transmission. 

\begin{figure}[t!]
  \centering
  \begin{subfigure}{0.49\columnwidth}
    \centering
    \includegraphics[width=\linewidth]{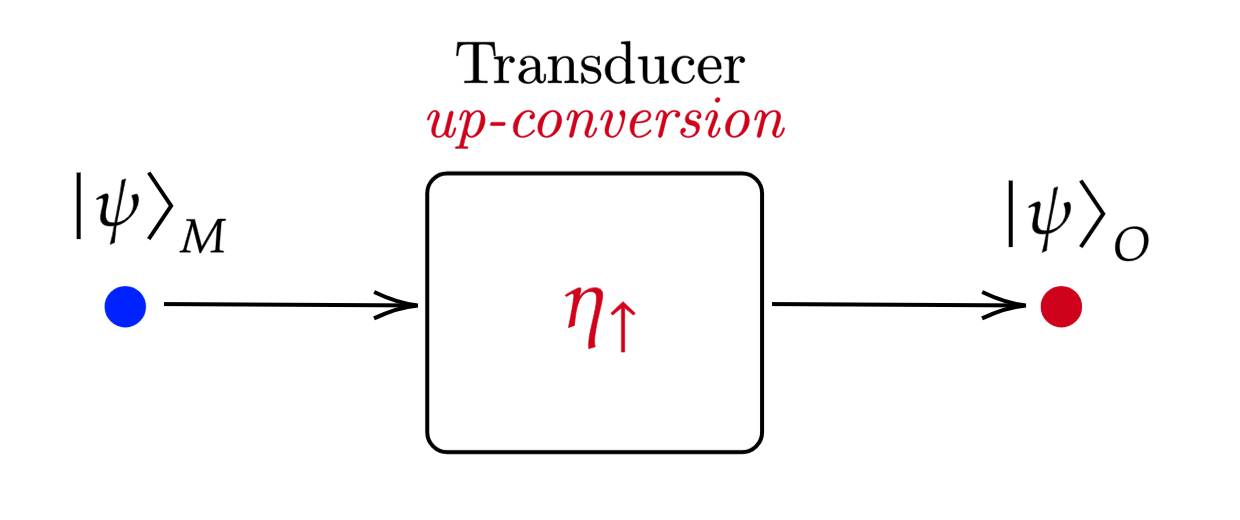}
    \caption{}
    \label{fig:02.1}
  \end{subfigure}
  \hfill
  \begin{subfigure}{0.49\columnwidth}
    \centering
    \includegraphics[width=\linewidth]{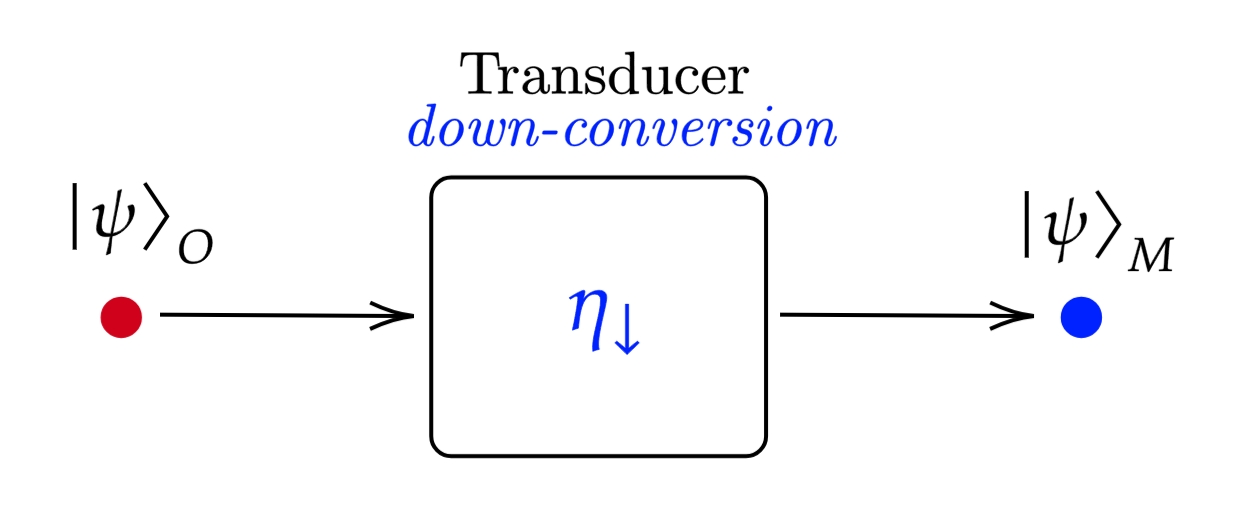}
    \caption{}
    \label{fig:02.2}
  \end{subfigure}
  
  \vspace{0.4cm} 
  \begin{subfigure}{\columnwidth}
    \centering
    \includegraphics[width=\linewidth]{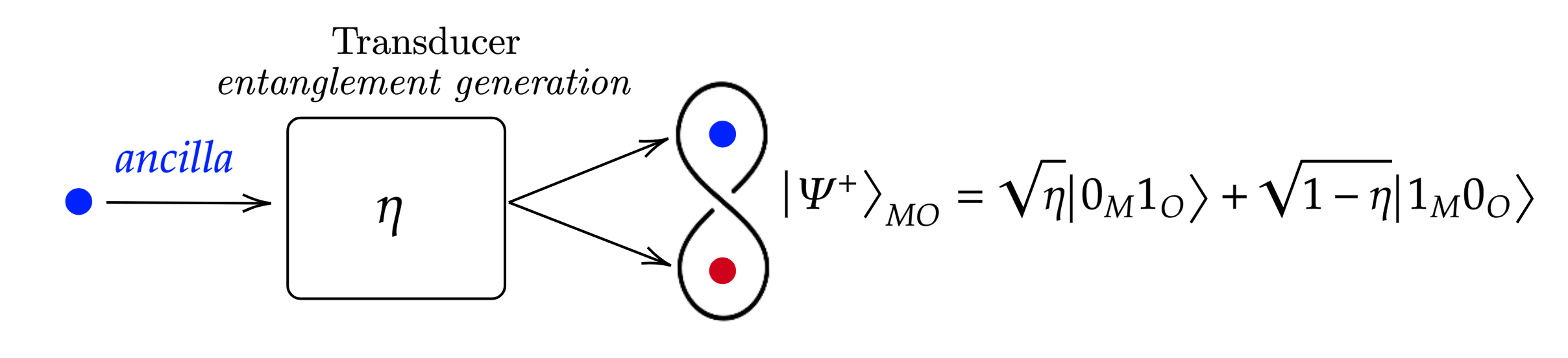}
    \caption{}
    \label{fig:02.3}
  \end{subfigure}

  \caption{Different functionalities of a quantum transducer: (a) up-conversion from microwave to optical, (b) down-conversion from optical to microwave, and (c) hybrid EPR pair generation. Subscripts of the quantum state indicate the frequency of the photon.}
\hrulefill

  \label{fig:tre_fig}
\end{figure}

\subsection{EQT: Strategy Description}
\label{sec:02.2}
As mentioned above, the EQT strategy exploits quantum transduction for hybrid EPR generation, instead of converting the information qubit from one frequency to another as done in DQT.
In other words, while in DQT a quantum transducer is responsible for a frequency conversion of the quantum information to be transmitted, in EQT the transducer generates entanglement for quantum teleportation.
While the state-of-the-art technology does not enable the achievement of high values of conversion efficiency, it is instead possible to generate \textit{hybrid} entanglement, i.e., entanglement between two different (optical and microwave) domains \cite{Tsa-11, WuCuiFan-21}. In the proposed EQT strategy, hybrid entanglement is generated at both source and destination,
therefore the entanglement generation occurs ``at both ends'' \cite{CacCalVan-20,KozWehVan-22}.
Specifically, two different interactions can lead to the generation of entanglement within a quantum transducer. On the one hand, a spontaneous parametric down-conversion (SPDC) of an input pump field (\textit{blue detuned pump})  
can generate entanglement between optical and microwave fields within the transducer \cite{WuCuiFan-21, KraRanHol-21}. 
SPDC is a non-linear optical process where a photon spontaneously splits into two photons of lower energies \cite{Cou-18}.
On the other hand, with a specific initialization of a microwave field inside the transducer, a beam splitter interaction enabled by a different frequency input pump (\textit{red detuning}) can lead to the conversion of the initialized microwave photon into an optical photon \cite{KraRanHol-21, Han-25}, as schematically depicted in Fig.~\ref{fig:02.3}.
In this second scenario, the generated entangled state can be expressed as the following \textit{Fock state}:
\begin{align}
    \label{eq:01}
    \ket{\Psi^s_{MO}}=\frac{1}{\sqrt{2}}(\ket{0_{M}^s1_{O}^s}+\ket{1_{M}^{s}0_{O}^{s}}).
\end{align}
with the subscripts $(\cdot_M)$ and $(\cdot_O)$ denoting the photon domain (i.e., microwave and optical), and the superscript $(\cdot^s)$ denoting the ``location'' of each ebit at the source.
Specifically, in \eqref{eq:01} the term $\ket{0_{M}^s 1_{O}^s}$ denotes the event when the microwave photon is successfully converted into an optical photon, while the term $\ket{1_{M}^s 0_{O}^s}$ denotes the event when the microwave photon is not converted into an optical photon. 
\begin{remark}
    The assumption of an EPR state -- i.e., a maximally entangled state in \eqref{eq:01} -- depends on a careful setting of the transduction hardware parameters \cite{KraRanHol-21}. For instance, in the case of beam splitter interaction, having $\eta_{\uparrow}=0.5$ is a necessary condition for maximizing the entanglement \cite{DavCacCal-24-1}.
\end{remark}

\begin{remark}
In the case of SPDC, the generated hybrid entanglement is in the form $\ket{\Phi^s_{M,O}} \approx \sqrt{1-|\lambda|^2}(\ket{0^s_{M}0^s_{O}}+\lambda \ket{1^s_{M}1^s_{O}})$, where $\lambda$ is the effective squeezing factor and $|\lambda|^2$ represents the probability of a couple of photon generation \cite{ZhoWanZou-20}.
%which is equivalent to \eqref{eq:01} up to a basis change. 
However, in the network simulator, we model the transducer as a component able to perform beam splitter interaction, as described in Sec.\ref{sec:03}. 
With this strategy, a single-component model can simulate both direct conversion and hybrid entanglement generation.
\end{remark}

Once the entanglement generation process is performed within the source and destination transducers, the optical photons of each generated EPR are transmitted through optical quantum channels to a beam splitter in the middle of the link followed by two detectors. 
This setup is unable to distinguish the \textit{which-path} information \cite{KraRanHol-21, PakZanTav-17}. When one of the two detectors clicks, it indicates that at least one optical photon has been generated. However, due to path erasure, we cannot determine whether this photon was generated by conversion from the microwave-initialized state at the source or at the destination. As a result, we are unable to determine if there is a microwave photon present at either location. 
This generates a \textit{path-entanglement} \cite{MontVivCap-15} between the microwave photon at the source and the one at the destination \cite{KraRanHol-21}, thus the overall effect of the beam splitter and detectors performs entanglement swapping \cite{BrieDurCir-98}.
Specifically, the detectors project the received optical photons into a Bell state and the heralded signal (i.e., the detector-click) indicates the distribution of entanglement between the remote superconducting processors \cite{DuaLukCir-01}:
\begin{align}
    \label{eq:02}
    \ket{\Psi_{MM}^{s,d}}=\frac{1}{\sqrt{2}}(\ket{0_{M}^s1_{M}^d}+\ket{1_{M}^s0_{M}^d}).
\end{align} 
Once entanglement between source $s$ and the destination $d$ nodes is heralded, the teleportation of the information qubit can now be performed.
One of the advantages of EQT over DQT is the fact that the quantum information to be transferred is never directly converted from one frequency to another. On the contrary, the frequency conversion acts on the entanglement only, which, being a communication resource rather than information, it is not constrained by the no-cloning theorem \cite{IllCalMan-22}. Thus, even if the photon encoding the quantum correlation is lost during transmission through optical channels, it can be re-transmitted multiple times, as many as needed so entanglement is generated.
\begin{remark}
    The proposed scheme assumes that
    microwave photon conversion into optical is successful in only one end (either source or destination). On the contrary, it may happen that both conversion processes succeed, resulting in two emitted optical photons arriving at the beam-splitter and detectors setup. As a result, the state shared between source and destination is $\ket{0^s_{M}}\ket{0^d_{M}}$ and it is not the entangled state in Eq.~\eqref{eq:02}.
    However, due to path erasure, if the detectors used are not photon-number resolving, one detector click is triggered despite two photons reaching the beam-splitter and detectors setup, resulting in the erroneous heralding of entanglement between the remote nodes. 
\end{remark}

\section{SeQUeNCe Module Design}
\label{sec:03}

In this section, we show our design and implementation of the direct quantum transduction (DQT) and entanglement-based quantum transduction (EQT) strategies in SeQUeNCe,
a customizable discrete event simulator of quantum networks~\cite{WuKolChu-21}.
The software framework of SeQUeNCe abstracts a quantum network architecture composed of several modules.
Among them, the \textit{hardware module} is used to model elementary hardware building blocks of quantum networks including quantum gates, quantum memories, quantum channels, and classical channels. 
Our contribution extends the hardware module of SeQUeNCe with the first quantum transducer component that \textit{models the conversion of a microwave photon into an optical photon and vice-versa via a beam splitter interaction}.
As mentioned in Sec.~\ref{sec:DQT}, the conversion probability is determined by the system's conversion efficiency. Therefore we set the conversion efficiency as an adjustable parameter in our transducer component.
Specifically, the transducer component has one input that can be microwave or optical depending on the ``direction'' of the desired conversion, and two outputs, one optical and one microwave, indicating whether the direct conversion is successful or not.
While a quantum transducer can have different physical implementations \cite{LauSinBar-20} (e.g., opto-electro-mechanical and electro-optical as anticipated in Sec.\ref{sec:02}), the proposed model is independent from the physical realization of the transducer, allowing the proposed study to be abstracted from specific hardware.
Moreover, as anticipated in the remark of Sec.~\ref{sec:02.2}, the same component is exploited to model both the direct conversions of the quantum information to be transferred in the DQT strategy and the intrinsic path-entanglement generation to be distributed in the EQT strategy.
\begin{remark}
    Obviously, the quantum transduction problem does not reduce to a probabilistic frequency conversion. Interfacing between several different hardware platforms presents several challenges, such as mode-mismatching, which includes conditions of impedance, spatial overlap, and the temporal properties of microwave and optical signals \cite{RueSedCol-16}.
    However, we can embed these requirements in the conversion efficiency parameter, allowing us to obtain in SeQUeNCe a software component that is easy to configure for the required analyses and easy to scale for large simulations.
\end{remark}

Besides the implementation of the transducer, other custom components have been created within the hardware module of SeQUeNCe to implement the proposed strategies, as we further discuss in the next sections.

\subsection{DQT Module Design}
\label{sec:3.1}

\begin{figure}[t!]
    \centering
    \includegraphics[width=1\linewidth]{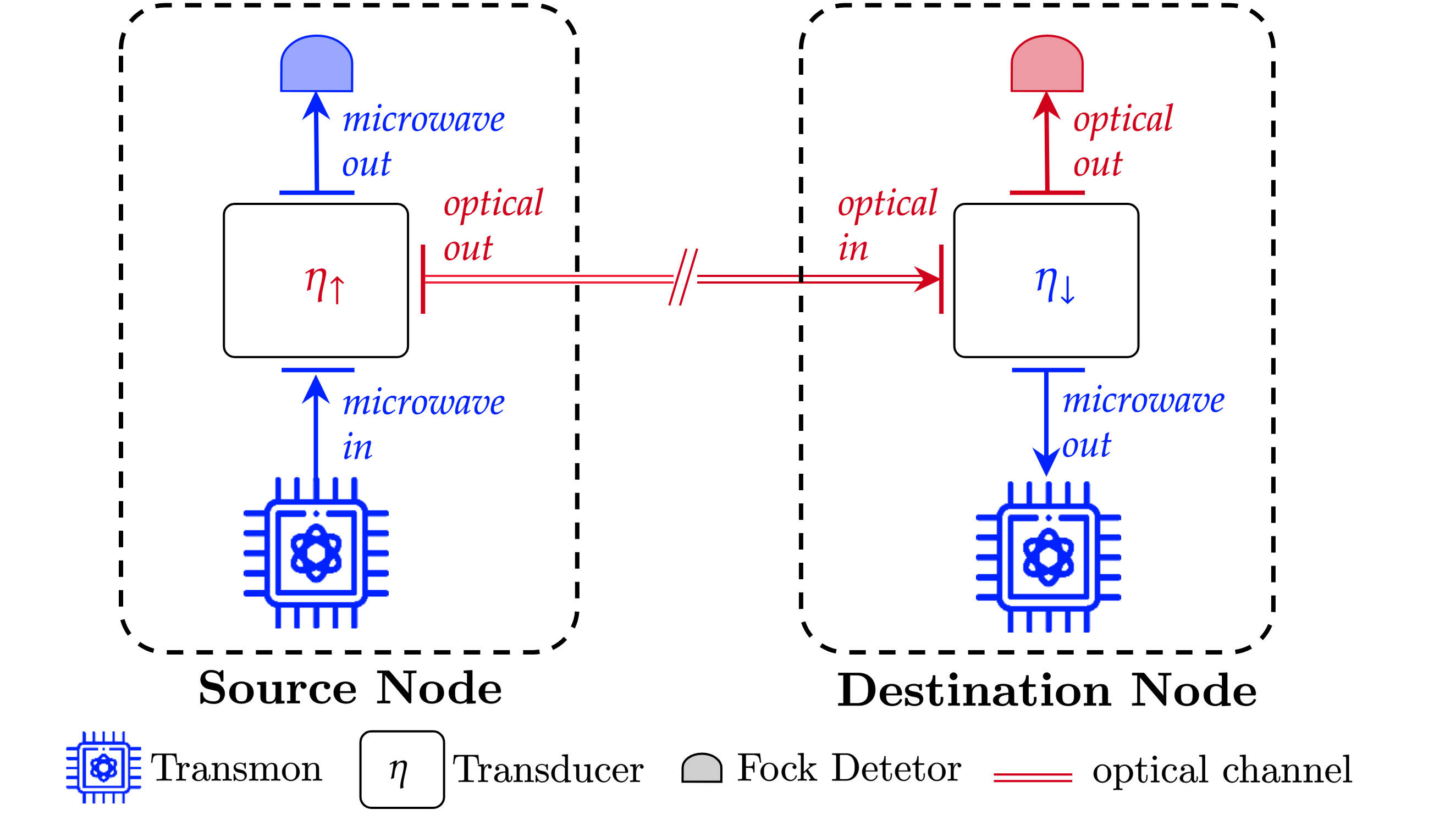}
    \caption{DQT system setup.}
    \hrulefill
    \label{fig:03}
\end{figure}

\subsubsection{Set-up}

DQT strategy simulations aim to evaluate the probability of successful distribution of quantum information. Specifically, at the source node, a transmon -- an example of superconducting qubit implementation -- stores the quantum information to be transmitted, and it emits microwave photons to the transducer within the source node. Upon receiving a microwave photon, the transducer may or may not convert it into an optical photon.
If the conversion fails, the unconverted microwave photon is detected by the click of a microwave detector within the source node.
If the conversion is successful, microwave photons successfully converted to the optical domain are sent via optical fiber to the destination node where a second transducer can perform a down-conversion.
At the destination node, if the down-conversion fails, an optical detector within the receiver node is triggered. If the down-conversion is successful, the microwave photon is sent to the destination transmon that can successfully update its quantum state.  Fig.~\ref{fig:03} shows the system setup.

\subsubsection{Design}

\begin{figure}[t!]
    \centering
    
    \includegraphics[width=1\linewidth]{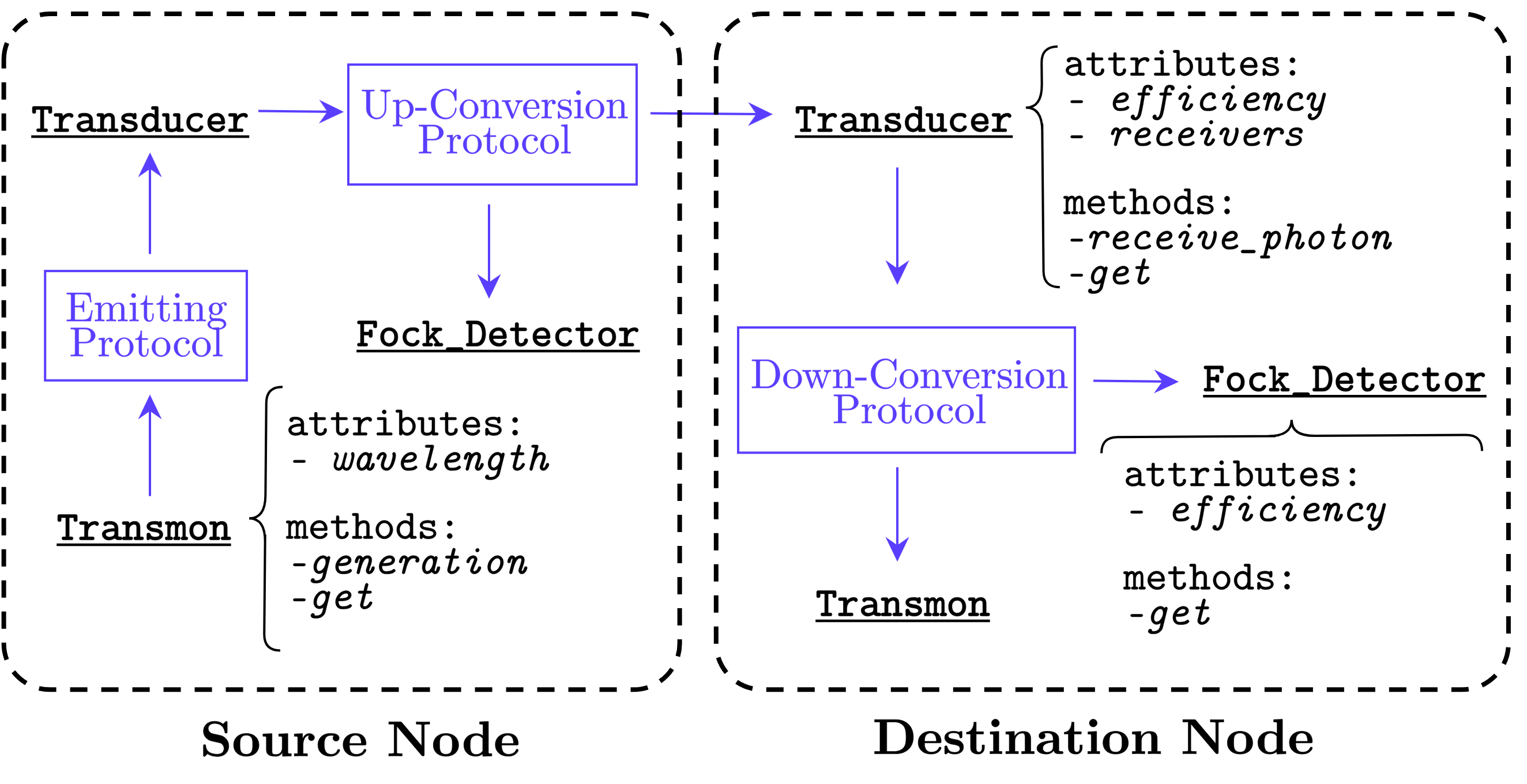}
    \caption{DQT high-level design. Attributes can be manually set while methods are called within the protocols.}
    \hrulefill
    \label{fig:04}
\end{figure}
For the proposed point-to-point DQT communication scheme in SeQUeNCe, both Source and Destination nodes are inherited from the \texttt{Node} class, while the quantum channel is inherited from the \texttt{OpticalChannel} class. 
As depicted in Fig.~\ref{fig:03}, each node has three main hardware components: transmon, transducer, and Fock\_detector. 
The hardware components have the following attributes and methods:

\begin{itemize}
    \item \texttt{Transmon}
        \begin{itemize}
            \item \textit{generation method}: generates a microwave photon,
            \item \textit{wavelength attribute}: frequency of the output microwave photon,
            \item \textit{get method}: keeps track of the received microwave photons.
        \end{itemize} 
    
    \item \texttt{Transducer}: 
        \begin{itemize}
            \item \textit{efficiency attribute}: sets the bi-directional transducer conversion efficiency,
            \item \textit{receivers attribute}: list of the transducer outputs,
            \item \textit{receive\_photon method}: keeps track of the received photons from a transmon and starts an up-conversion,
            \item
            \textit{get method}: keeps track of the received photons from a quantum channel and starts a down-conversion.
        \end{itemize} 
    \item \texttt{Fock\_Detector}: 
        \begin{itemize}
            \item \textit{efficiency attribute}: sets the detector efficiency,
            \item \textit{get method}: keeps track of received photons.
        \end{itemize} 
\end{itemize}
The class structure of our implementation enables a modular approach that allows individual, functional components to be easily reused for other case studies.
Attributes of each component have to be manually set within the software.
After the custom components are initialized, the custom protocols are created to control and monitor hardware: \textit{Emitting protocol},  \textit{Up-Conversion protocol}, and \textit{Down-conversion protocol}.

Specifically:
\begin{itemize}
    \item \textit{Emitting protocol}: calls for the generation method of a transmon component and sends the microwave-generated photon to the transmon receiver,
    \item \textit{Up-conversion protocol}: converts the generated microwave photon at the transducer into an optical photon. Then the optical photon is sent to the optical channel that has to transmit the photon to the receiver node,
    \item \textit{Down-conversion protocol}: converts the optical photon that reaches the transducer into a microwave photon, then sends it to the transducer receiver.
\end{itemize}
A high-level block diagram in Fig.~\ref{fig:04} summarizes our custom components, highlighting some of their key attributes and methods, and the protocols implemented.

\subsection{EQT Module Design}
\label{sec:03.2}

\begin{figure}[t!]
    \centering
   
    \includegraphics[width=\linewidth]{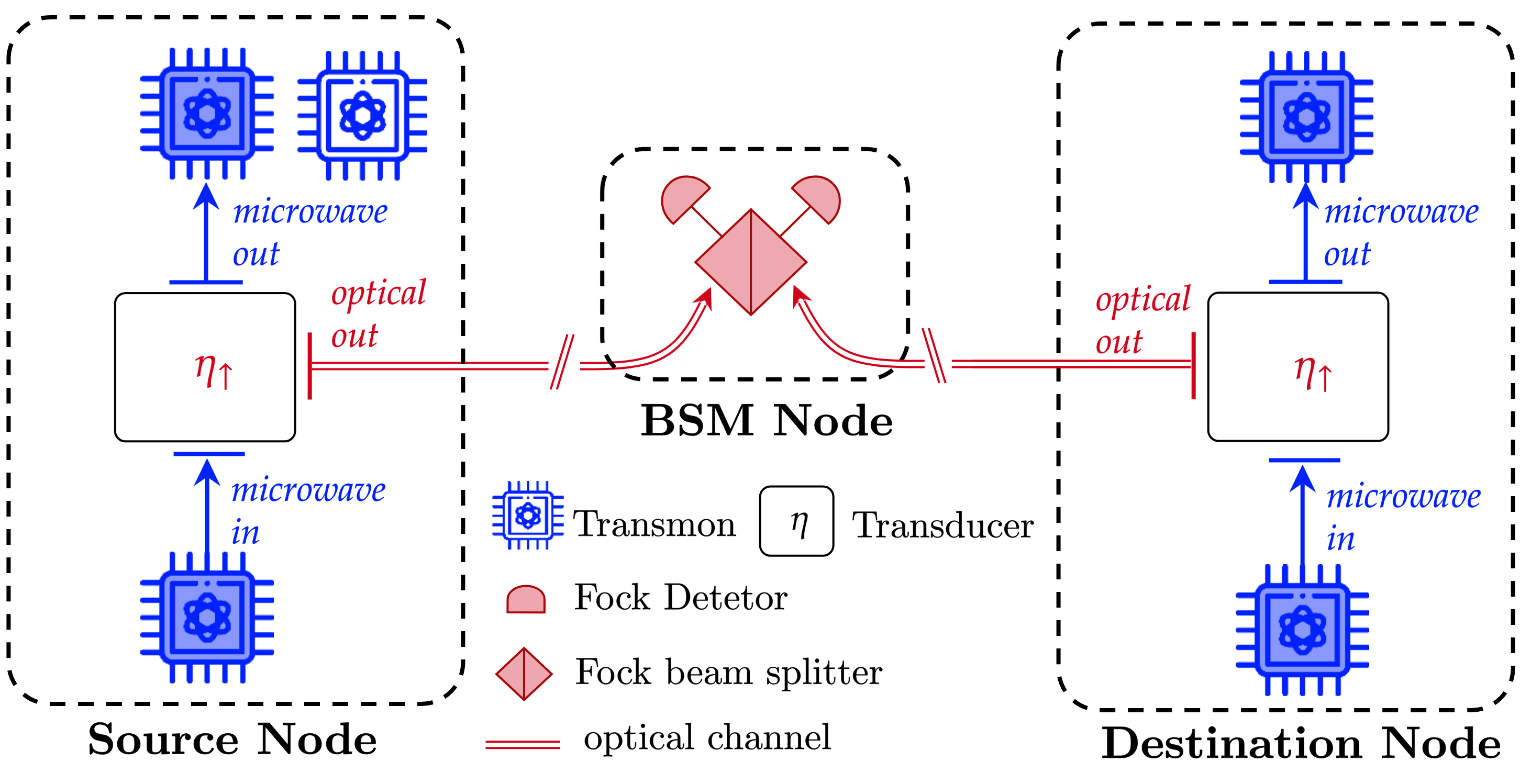}
    \caption{EQT system setup. Shaded blue transmons represent the ancilla qubits used for hybrid entanglement generation, while the non-shaded transmon represents the qubit that stores the quantum information to be teleported. Classical channels are omitted to maintain clarity.}
    \hrulefill
    \label{fig:05}
\end{figure}

\subsubsection{Setup}
For the EQT strategy, entanglement generation and distribution is the most complex process that requires a detailed discussion.
Therefore, in SeQUeNCe we study the entanglement distribution process and evaluate the percentage of successfully entangled pairs distributed, assuming that the subsequent teleportation protocol is noiseless. In other words, we can assume that if the entanglement is successfully generated and distributed the quantum information itself has been successfully transmitted from source to destination because of zero-noise local operation and classical communication (LOCC).
Specifically, in the EQT strategy microwave photons are sent from two transmons -- one at the Source Node and one at the Destination Node -- to their respective transducers.
It is important to highlight that, differently from DQT, in EQT neither of the microwave photons sent by the transmons constitute the quantum information to be transmitted, but they are ``ancilla'' qubits that generate hybrid entanglement through transduction (depicted in shaded blue in Fig.~\ref{fig:05}).
After receiving the microwave photons, both transducers implement an up-conversion process sending the eventually converted optical photons to a Bell state measurement (BSM) node that performs the entanglement swapping as explained in detail in Sec. \ref{sec:02.2}.
Fig.~\ref{fig:05} shows the system setup, while Fig.~\ref{fig:06} shows the high-level design of the components in SeQUeNCe.
Differently from DQT, in EQT if the up-conversion fails, the non-converted microwave photons are sent to another transmon on each node rather than to microwave detectors. This is done for two key reasons:
(1) a detection of the microwave photon measures the quantum state of the qubit, 
effectively destroying the entanglement itself; and
(2) after the entanglement distribution process succeeds, the transmon at the source that receives or not the microwave photon has to interact with the transmon (the non-shaded transmon in Fig.~\ref{fig:05}) that stores the quantum information to perform quantum teleportation.

\begin{figure}[t!]
    \centering
    
    \includegraphics[width=1\linewidth]{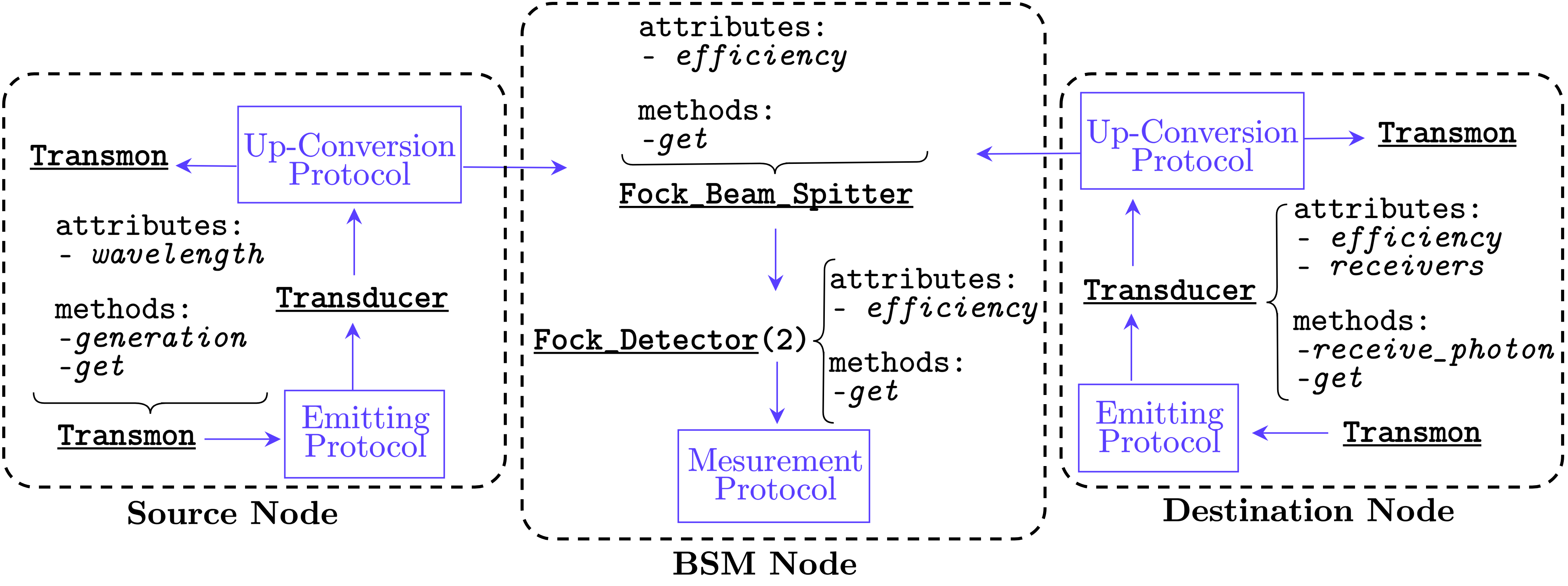}
    \caption{EQT high-level design. Attributes can be manually set while methods are called within the protocols.}
    \hrulefill
    \label{fig:06}
\end{figure}

%\begin{figure*}[t!]
%	\centering
 %   \begin{minipage}[c]{.32\linewidth}
	%	\centering
	%	\includegraphics[width=1\columnwidth]{Figures/Fig05.1.png}
		%\subcaption{$\eta_{\uparrow}^s=\eta_{\downarrow}^d=0.8$ }
	%	\label{fig:05.1}
	%\end{minipage}
	%\begin{minipage}[c]{.32\linewidth}
	%	\centering
	%	\includegraphics[width=1\columnwidth]{Figures/Fig05.2.png}
		%\subcaption{$\eta_{\uparrow}^s=\eta_{\downarrow}^d=0.5$ }
		%\label{fig:05.2}
	%\end{minipage}
    %	\begin{minipage}[c]{.32\linewidth}
	%	\centering
	%	\includegraphics[width=1\columnwidth]{Figures/Fig05.3.png}
		%\subcaption{$\eta_{\uparrow}^s=\eta_{\downarrow}^d=0.1$}
		%\label{fig:05.3}
%	\end{minipage}

%	\hrulefill
%\end{figure*}

\subsubsection{Design}

In EQT, each node of the setup is customized with different
hardware components.
Besides the components presented in Sec.~\ref{sec:3.1}, \texttt{Fock\_Beam\_splitter} is additionally introduced:

\texttt{Fock\_Beam\_splitter}:
\begin{itemize} [label=-]
    \item \textit{receivers attributes}: output ports of the component,
    \item  \textit{get method}: sends optical photons into one of the two receivers.
    If two indistinguishable optical photons reach the beam splitter, they will both exit through the same output port, according to the Hong-Ou-Mandel interface \cite{Bra-17}.
\end{itemize}

The custom protocols created to perform EQT are \textit{Emitting protocol} and
\textit{Up-Conversion protocol}, as in DQT. Besides, the \textit{Measurement protocol} is created:
\begin{itemize}
    \item \textit{Measurement protocol}: keeps track of the received optical photons by the Fock detectors and updates the counts according to the type of detectors used.
\end{itemize}

\begin{figure}[t!]
    \vspace{-0.4cm} 
    \centering
	\begin{minipage}[c]{0.98\linewidth}
		\centering
		\includegraphics[width=1\columnwidth]{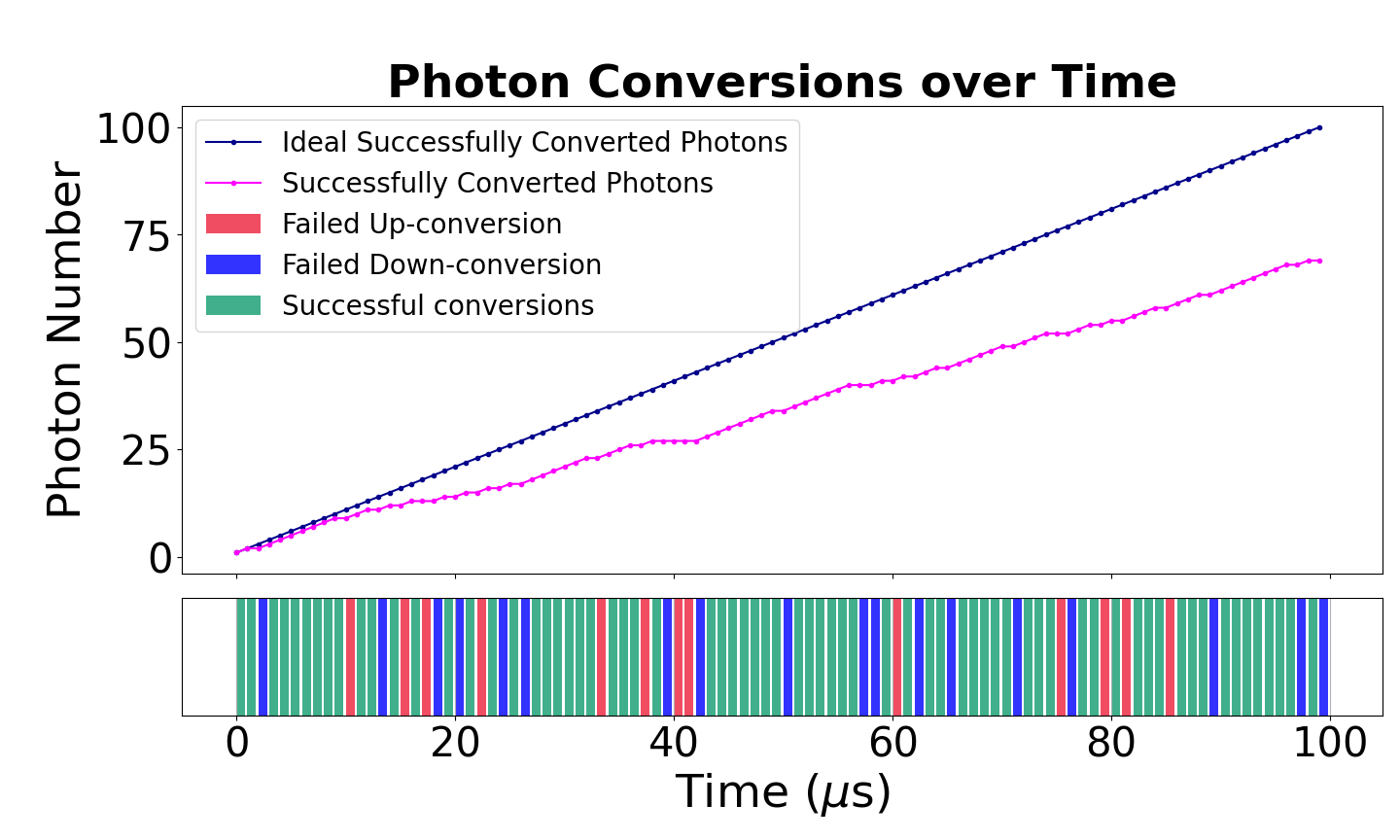}
            \vspace{-0.2in}
		\subcaption{$\eta_{\uparrow}^s=\eta_{\downarrow}^d=0.8$}
		\label{fig:07.1}
	\end{minipage}

        \vspace{0.1cm} 

	%\hfill\hspace{3pt}
	\begin{minipage}[c]{0.98\linewidth}
		\centering
		\includegraphics[width=1\columnwidth]{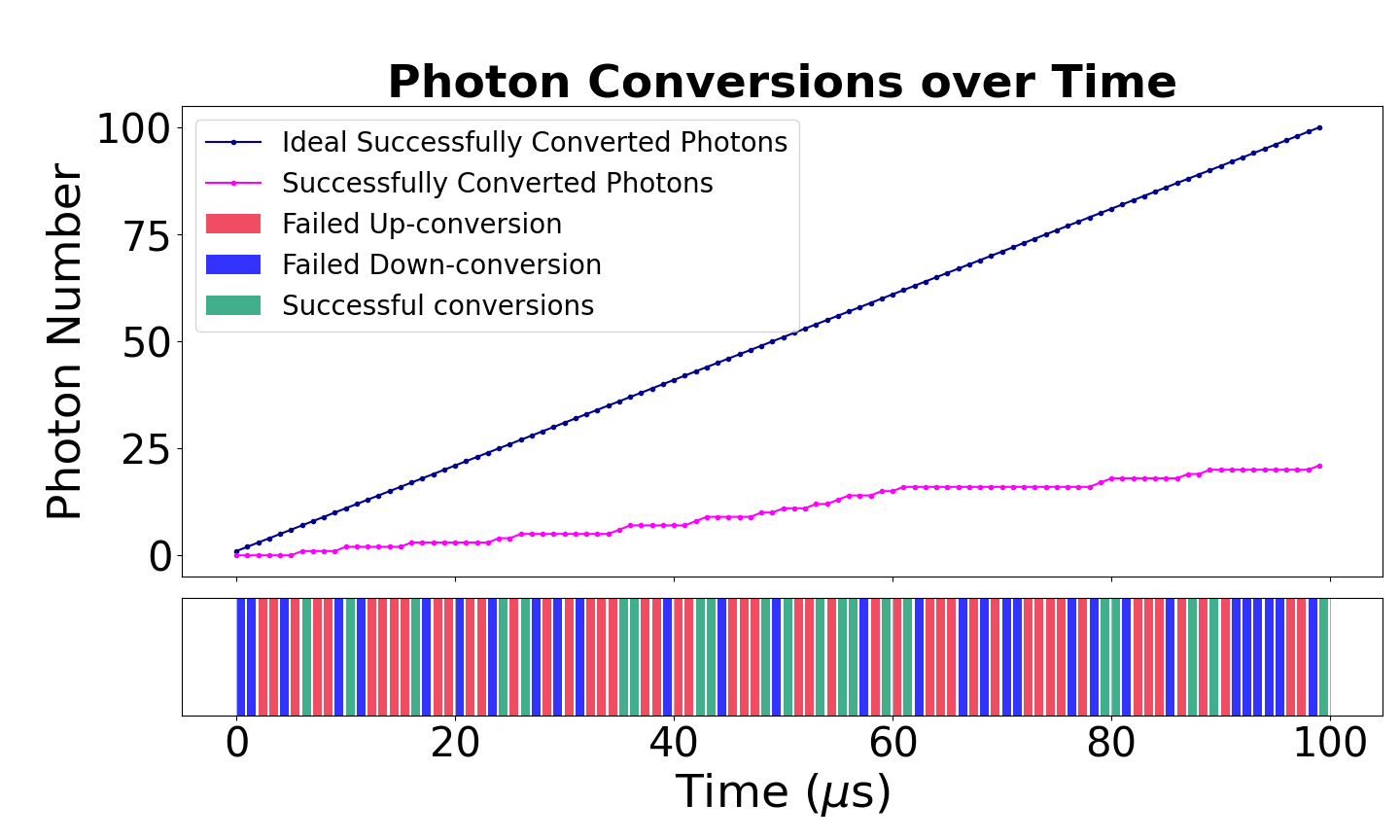}
            \vspace{-0.2in}
		\subcaption{$\eta_{\uparrow}^s=\eta_{\downarrow}^d=0.5$}
		\label{fig:07.2}
	\end{minipage}

        \vspace{0.1cm} 

    \begin{minipage}[c]{0.98\linewidth}
		\centering
		\includegraphics[width=1\columnwidth]{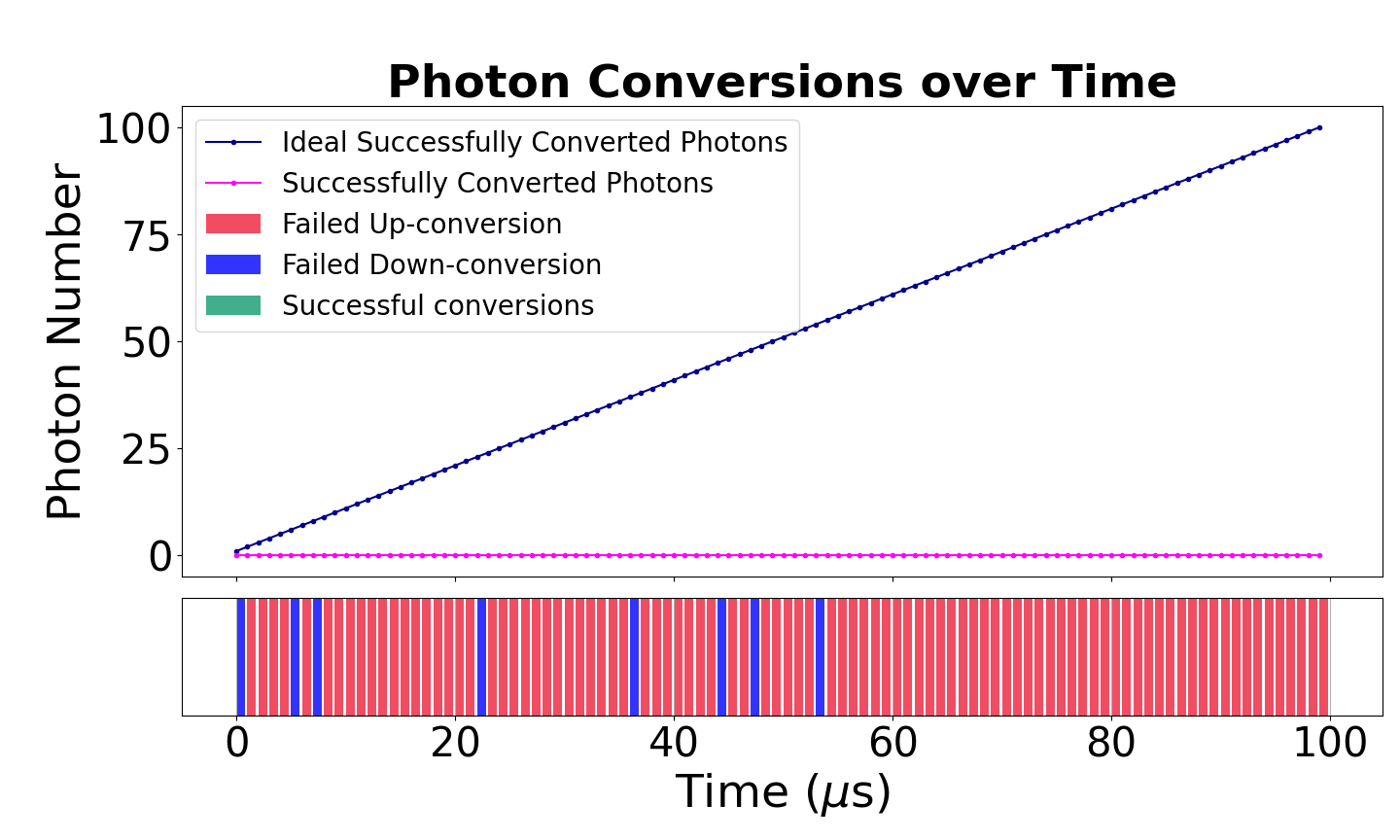}
            \vspace{-0.2in}
		\subcaption{$\eta_{\uparrow}^s=\eta_{\downarrow}^d=0.1$}
		\label{fig:07.3}
	\end{minipage}

	\caption{SeQUeNCe simulations results of DQT strategy with $l_{s,d}<<L_0$.}
	\label{fig:07}
	\hrulefill
\end{figure}

\section{Simulation Results}
\label{sec:04}

To compare the proposed strategies, we evaluate some communication metrics through numerical simulations in SeQUeNCe. We consider different scenarios where the transducers involved in the setup have different values of conversion efficiency.

%two scenarios. The first one where the transducers involved in the setup are high efficiency (for example 80\% efficiency), in the second one with 50\% efficiency is considered (reached experimentally in \cite{AndPetPur-14, HigBurUrm-18}).

\subsection{DQT Simulation Results}
\label{sec:04.1}
We first consider DQT strategy.
The probability of a successful quantum information transfer $p$ %$p=\frac{n_s}{N}$ 
in a point-to-point link can be expressed as a function of the transducer conversion efficiency as follows \cite{DavCacCal-24-1}:
\begin{equation}
    \label{eq:03}
    p=\eta_\uparrow^s\eta_\downarrow^de^{-\frac{l_{s,d}}{L_0}}
\end{equation}
where the superscripts $(\cdot^s)$ and $(\cdot^d)$ denote the ``location'' of the transducer at the source and destination nodes, respectively.
The term $e^{-\frac{l_{s,d}}{L_0}}$ %constitutes the efficiency of the quantum channel 
takes into account the decay effects of the channel%on the qubit
, where $l_{s,d}$ is the length of the fiber link between source and destination and $L_0$ is the attenuation length of the fiber. As an instance, optical photons with a wavelength equal to $1550$nm -- experience an attenuation of $0.2$~dB/km in commercial optical fibers, i.e. $L_0=22$~km\cite{SanChrder-11}.
%By substituting the values of conversion efficiencies and fiber characteristics in Eq.~\ref{eq:03}, we obtainthe \textit{expected value} of $p$. 

%Specifically, in order to distinguish the values of $p$ obtained by the theoretical analysis and the one obtained by simulation results,

Let us now evaluate $p$ with SeQUeNCe, where the emission of a microwave photon from the transmon at the source is periodically simulated. The simulation period is approximately $1 \mu s$, which constitutes the reset time of the microwave cavity in the quantum transducer \cite{KraRanHol-21, KreDovLiv-16}.
We consider only the microwave cavity reset time, since it is orders of magnitude longer than the detection times at both source and destination nodes (typically around few $ps$ \cite{Had-12}).
We can model both conversions as Bernullian variables, where the probability of successful conversion is given by the conversion efficiency of the transducer. The number of trials $N$ we perform in SeQUeNCe is fixed for all the simulations at $N=100$.

\begin{table}[]
    \centering
    \normalsize
    \caption{DQT probabilities of successful quantum information transmission for different values of conversion efficiencies.}
    \begin{tabular}{|c|c|c|}
    \hline
         $\eta_\uparrow^s=\eta_\downarrow^d$ & $p$  & $p^s$ \\
         \hline
         0.8 & 0.72 & 0.69\\
         \hline
        0.5 & 0.25 & 0.21\\
        \hline
         0.1 & 0.01 &  0\\
         \hline
    \end{tabular}
    \label{tab:01}
\end{table}

%is determined by the chosen standard error $SE$ of the product of the variables, set at $5\%$.
In our simulation results the probability of a successful qubit transmission is given by the simulated ($n_s$) vs. the ideal ($n_i$) number of photons successfully transmitted, where $n_i$ implies lossless transducers ($\eta_\uparrow^s=\eta_\downarrow^d=1$), i.e., one successfully transmitted photon for each period ($n_i=N$). We call the probability of successful transmission obtained with SeQUeNCe the \textit{simulated probability}, denoted by $p^s$.
Fig.~\ref{fig:07} shows our simulation results presenting the number of converted photons over the simulation time and tracking when the up- and down-conversions have failed in each period.
Let us consider a first scenario where $\eta_{\uparrow}^s=\eta_{\downarrow}^d=0.80$ and $l_{s,d}<<L_0$. 
From \eqref{eq:03} it results $p=0.72$.
%\begin{figure}[t!]
%	\centering
%	\begin{minipage}[c]{\linewidth}
%		\centering
		%\includegraphics[width=1\columnwidth]{Figures/Fig05.1.png}
            %\vspace{-0.3in}
		%\subcaption{}
		%\label{fig:05.1}
	%\end{minipage}
	%\hfill\hspace{3pt}
	%\begin{minipage}[c]{\linewidth}
		%\centering
		%\includegraphics[width=1\columnwidth]{Figures/Fig05.2.png}
            %\vspace{-0.3in}
		%\subcaption{}
		%\label{fig:05.2}
	%\end{minipage}
	%\caption{Simulation of DQT with $\eta_{\uparrow}=\eta_{\downarrow}=0.85$ in (a) and $\eta_{\uparrow}=\eta_{\downarrow}=0.5$ in (b). 
    %}
	%\label{fig:05}
	%\hrulefill
%\end{figure}
%For instance, if we consider the case of $\eta_{\uparrow}^s=\eta_{\downarrow}^d=0.80$ and $l_{s,d}<<L_0$, from Eq.~\ref{eq:03} it results $p=0.72$. This constitutes the \textit{expected value} of the probability of successful qubit transfer.
Fig.~\ref{fig:07.1} shows the simulation results in these hypotheses, with $p^s=0.69$. 
Of course it is possible to obtain a closer approximation to the expected value $p=0.72$ from \eqref{eq:03} by increasing the number of microwave samples or, equivalently, extending the duration of the simulation.

%The simulation results show that it is possible to achieve $p=0.62$, a closer approximation to the expected value $p=0.72$ obtained from Eq.~\ref{eq:03},by increasing the number of microwave samples or, equivalently, extending the duration of the simulation.

A conversion efficiency of $80\%$ is well above values that can be achieved with the current state-of-the-art technology. %and a reduction of this value leads to a significant worsening of the communication performances of DQT strategy.
Therefore, let us now consider a second scenario, where the transducers' conversion efficiencies are set to $50\%$ (values reached experimentally for optomechanical devices in \cite{AndPetPur-14, HigBurUrm-18}) and again, $l_{s,d}<<L_0$.
In this case, \eqref{eq:03} predicts $p=0.25$, while from our SeQUeNCe simulation (showed in Fig.~\ref{fig:07.2}) we obtain $p^s=0.21$.
%results with transducers conversion efficiency set to $50\%$.
It is evident that the number of successfully transmitted microwave photons is lower than the previous case and the number of failed conversions dramatically increases. 
Indeed, a reduction of conversion efficiency value leads to a significant worsening of the communication performances of the DQT strategy.
%The maximum reachable value for $p$ is $0.25$ and the simulation results only give $p=0.14$

A conversion efficiency of $50\%$ is still a high value compared to the ones currently achievable with electro-optical transducers \cite{SahHeaRue-22}, which present attractive features from a communication perspective, such as low added noise in the conversion\cite{DavCacCal-24-1, LauSinBar-20}. 
Thus, we consider a third and last scenario, in which the conversion efficiency of both transducers in our simulation is set to $10\%$ (that constitutes an experimentally achieved value for electro-optical transducers \cite{SahHeaRue-22}).
In this case, from \eqref{eq:03} it results that $p$ just reaches $0.01$.
Fig.~\ref{fig:07.3} shows the simulation results in this hypothesis: none of the photons generated at the source reaches the destination, i.e. at least one conversion fails for each attempt ($p^s=0$).

\begin{figure}[t!]
    \vspace{-0.4cm} 
    \centering
	\begin{minipage}[c]{0.98\linewidth}
		\centering
		\includegraphics[width=1\columnwidth]{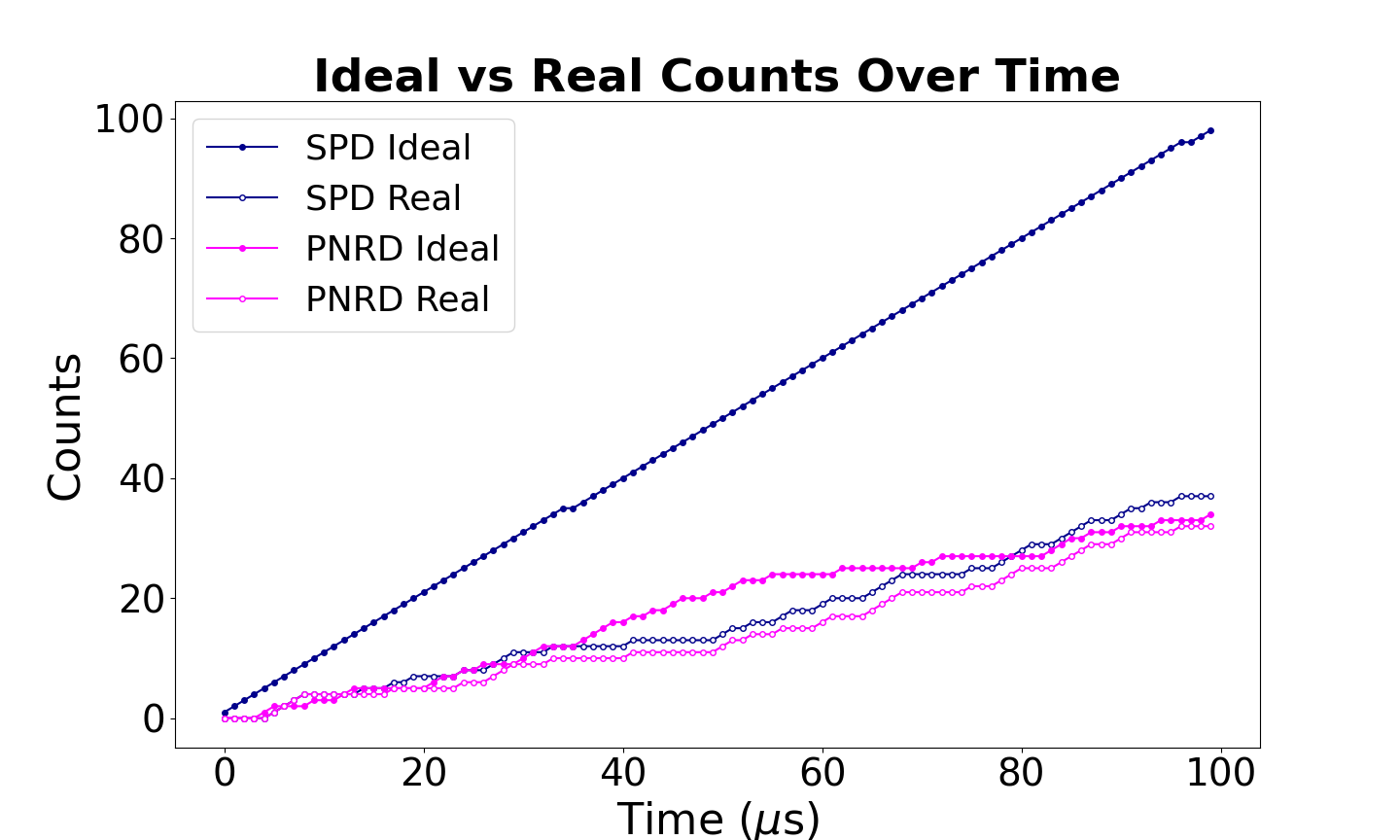}
            \vspace{-0.2in}
		\subcaption{$\eta_{\uparrow}^s=\eta_{\downarrow}^d=0.8$}
		\label{fig:08.1}
	\end{minipage}

        \vspace{0.1cm} 

	%\hfill\hspace{3pt}
	\begin{minipage}[c]{0.98\linewidth}
		\centering
		\includegraphics[width=1\columnwidth]{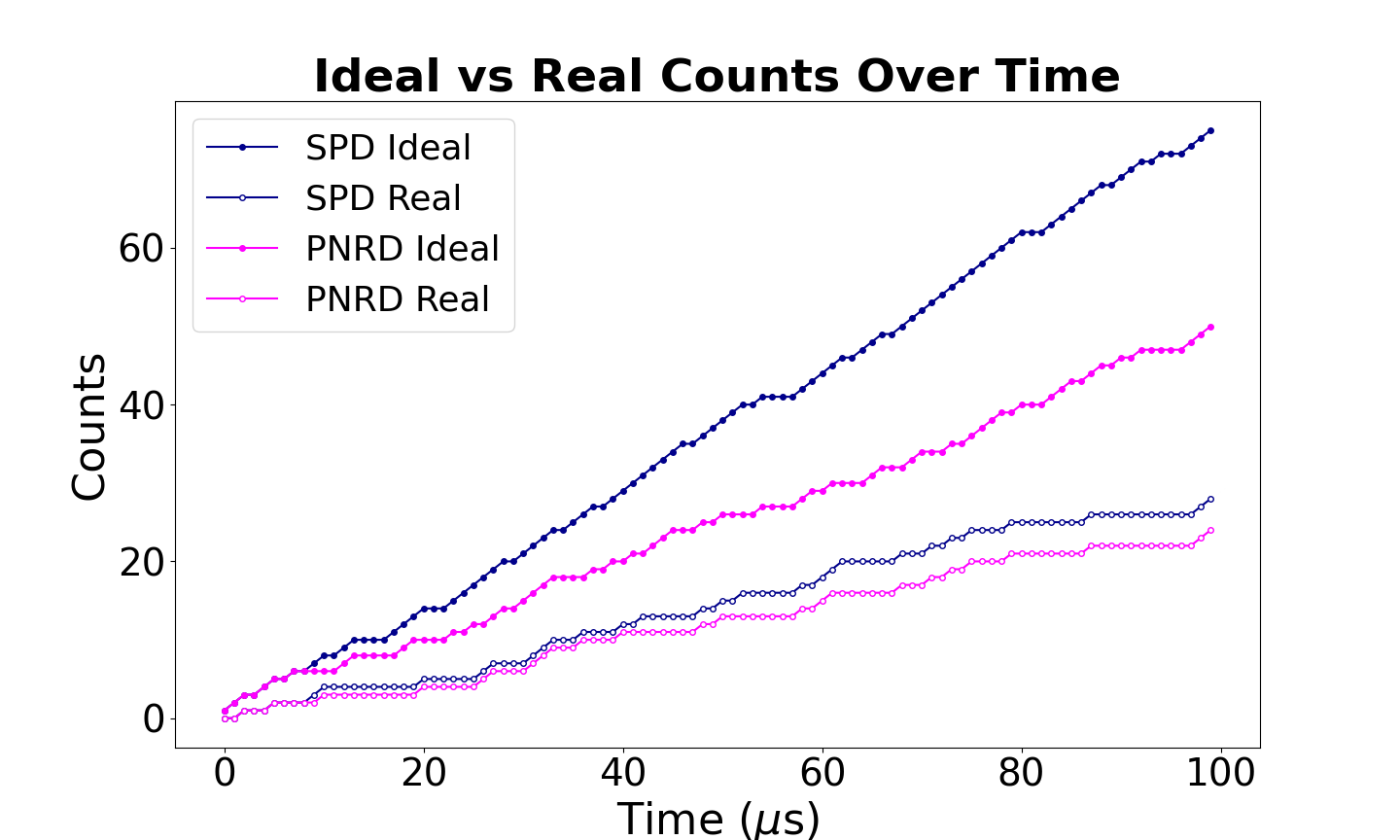}
            \vspace{-0.2in}
		\subcaption{$\eta_{\uparrow}^s=\eta_{\downarrow}^d=0.5$}
		\label{fig:08.2}
	\end{minipage}

        \vspace{0.1cm} 

    \begin{minipage}[c]{0.98\linewidth}
		\centering
		\includegraphics[width=1\columnwidth]{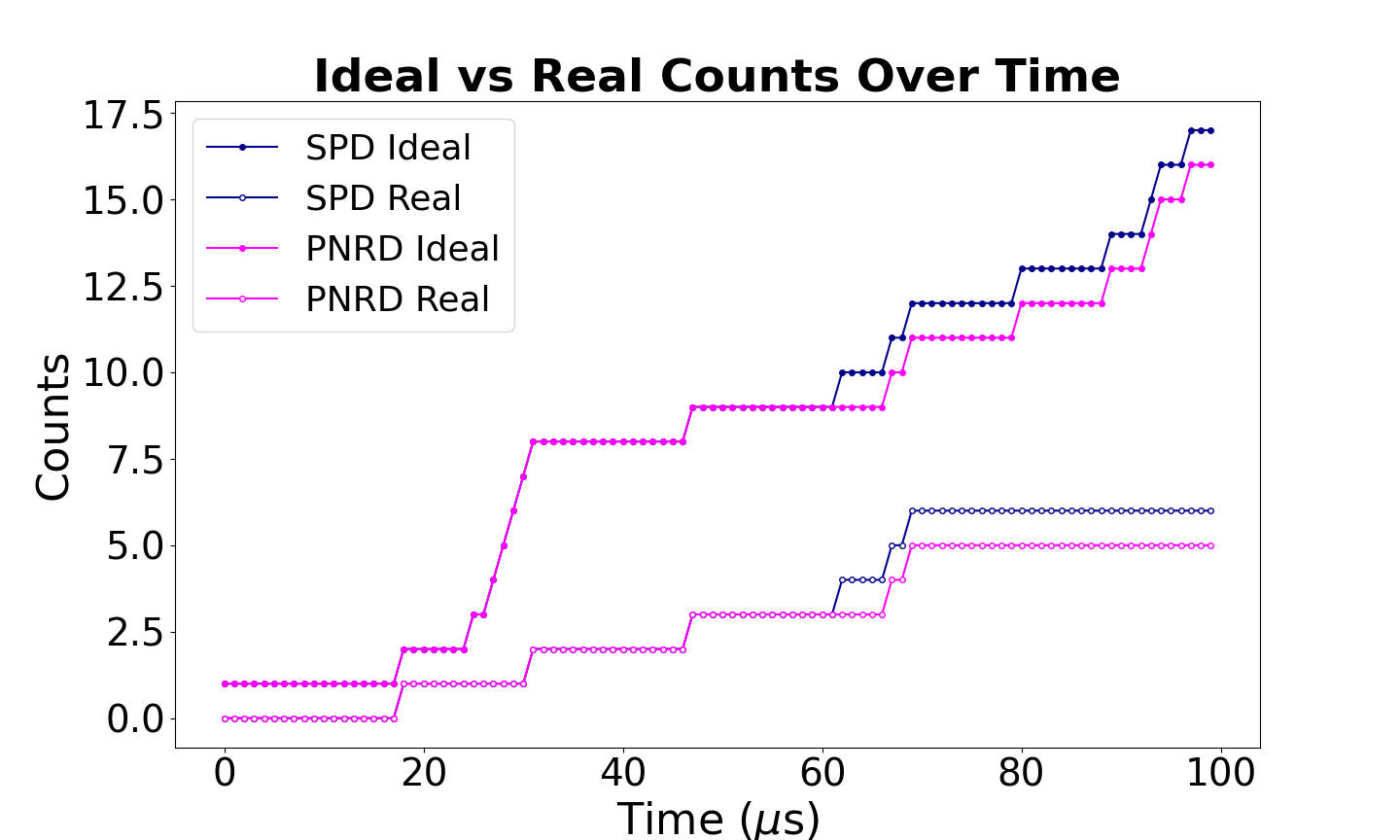}
            \vspace{-0.2in}
		\subcaption{$\eta_{\uparrow}^s=\eta_{\downarrow}^d=0.1$}
		\label{fig:08.3}
	\end{minipage}

	\caption{SeQUeNCe simulations results of EQT strategy with $l_{s,d}<<2L_0$.}
	\label{fig:08}
	\hrulefill
\end{figure}

Tab.~\ref{tab:01} summarizes the DQT simulation results for different values of conversion efficiency, with $l_{s,d}<<L_0$.

%\begin{remark}
   % It is important to notice that the number of trials $N$ (and as a consequence of the simulation time) for the different proposed SeQUeNCe simulations varies according to the conversion efficiency of the transducers within setup. Indeed, the choice of $N$ is dictated by the goal of obtaining the same standard  error $SE$ for all simulations.
%\end{remark}

\begin{remark}
        Let us notice that in our simulation results, the obtained $p^s$ are smaller than $p$ from \eqref{eq:03}. However, in general, the approximation obtained for a fixed value of $N$ may either overestimate or underestimate the value expected from \eqref{eq:03}.
        % was to analyze the different performances while keeping the standard error constant.
        These results demonstrate that our quantum transducer implementation in SeQUeNCe closely follows the theoretical models, thus it can be used on larger scale simulations of DQT.
\end{remark}

%The simulation results confirm that direct conversion is not a viable strategy for quantum information transmission. 

\subsection{EQT Simulation Results}
\label{sec:04.2}

In the case of EQT, the main parameter to characterize the performance of this communication strategy is the probability of EPR pair distribution between the remote nodes $p_e$. Indeed, as described in Sec.~\ref{sec:02.2}, once an EPR pair is distributed, quantum teleportation protocol can be performed for the transmission of quantum information.
Moreover, as anticipated in Sec.~\ref{sec:03.2}, for the sake of clarity, we assume that quantum teleportation is noiseless, i.e., zero noise LOCC. 
Indeed, in this hypothesis and assuming the entangled pair distributed between the remote nodes is an EPR pair (i.e., it is a maximally entangled state), it has been proved that the quantum teleportation is deterministic~\cite{FanLinZhu-03}. 
For instance, the impact of noisy LOCC on quantum teleportation has been analyzed in \cite{CacCalVan-20}, showing that noisy LOCC affect the fidelity of the teleported state.
Under these assumptions, the probability of EPR pair distribution $p_e$ coincides with the probability of transferring quantum information (defined as $p$ for DQT in Sec.~\ref{sec:04.1}) \cite{DavCacCal-24-1}. Moreover, with the proposed EQT strategy with entanglement swapping, the distributed entanglement is heralded at the BSM node. As a result, the probability of EPR pair distribution can coincide with the probability of a single detector click, which we call $p_c$.
%In EQT strategy, we call $p_e=\frac{n_s^e}{n_i^e}$ the probability of entanglement distribution. In general $p_e$ does not correspond to the probability of a detector click $p_c$ because of the noise introduced by the optical detectors. 
However, as anticipated in the remark of Sec.~\ref{sec:02.2} it could happen that the BSM node is reached by two optical photons. This means that the up-conversion processes at both transducers have been successful and there is no entanglement between source and destination nodes.
%In EQT protocol, we call $p_e=\frac{n_s^e}{n_i^e}$ the probability of entanglement distribution., where $n_s^e$ and $n_i^e$ constitute the simulated and ideal number of successfully distributed entangled photons, respectively; and $n_i^e$ coincide with the number of periods $N$.
%where $n_s^e$ and $n_i^e$ constitute the simulated and ideal number of successfully distributed entangled photons, respectively; and $n_i^e$ coincide with the number of periods $N$.
We can identify this specific case and rule it out as a heralded entanglement if at the BSM node we exploit  Photon Number Resolving Detectors (PNRDs) \cite{ProLukRa-20}. %are exploited. 
Indeed, PNRDs are able to distinguish if the detector trigger is given by one or more optical photons. Therefore, the case of two successful conversions is not identified as an entangled state shared between source and destination. 
In this scenario, the probability of a single click can be expressed as a function of the transducer conversion efficiency:
\begin{align}
    \label{eq:04}
    p_c^{PNRD}= 2 (\eta_\uparrow- \eta_\uparrow^2)e^{-\frac{l_{s,d}}{2L_0}}
\end{align}
Assuming that both transducers at source and destination have similar features, in \eqref{eq:04} it is unnecessary to distinguish between $\eta_{\uparrow}^s$ and $\eta_{\uparrow}^d$ \cite{DavCacCal-24-1}.
For the aforementioned considerations, in this scenario, it results that $p_c^{PNRD}=p_e$. 
And moreover, because of the hypothesis of deterministic quantum teleportation, it results that $p_c^{PNRD}=p$.
We can also take into account detectors with non-ideal detection efficiency $\eta_d$, given by the probability of registering a count if a photon arrives at the detector \cite{ProLukRa-20}. Equation \eqref{eq:04} is given by the hypothesis of ideal photon detectors ($\eta_d=1$).
Instead, if we consider optical detectors with  $\eta_d<1$, in \eqref{eq:04} 
every transducer conversion efficiency term has to be weighted
with $\eta_d$.  %are not ideal (i.e., the detector efficiency is lower than one) \textcolor{blue}{QUI}, 
As a result, in the case of non-ideal optical detectors, $p_c^{PNRD}$ can be lower or higher than $p_e$.
Indeed, lower efficiency can cause some entangled states to be lost or introduce erroneously herald entanglement and, therefore, dark counts.

Differently from PNRDs, Single Photon Detectors (SPDs) are not able to distinguish if a detector click is triggered by one or two photons \cite{Had-09}. In this scenario, the probability of a single detector click is expressed as follows:
\begin{align}
    \label{eq:05}
    p_c^{SPD}= (2 \eta_{\uparrow}- \eta_{\uparrow}^2)e^{-\frac{l_{s,d}}{2L_0}}
\end{align}
In this case, only a fraction $\frac{2 \eta_{\uparrow}- 2 \eta_{\uparrow}^2}{2 \eta_{\uparrow} - \eta_{\uparrow}^2}$ of clicks corresponds to entanglement generation, with the remaining click fraction corresponding to a failed attempt \cite{DavCacCal-24-1}. 
Furthermore, in the case of non-ideal SPDs (i.e., $\eta_d<1$), in \ref{eq:05} every transducer conversion efficiency term has to be weighted with $\eta_d$. This makes the detector count always lower than the case with ideal optical detectors.

Let us now evaluate $p_c$ for different detector types with SeQUeNCe. Again, we consider three different scenarios, varying the conversion efficiencies of the transducers involved in the setup and, similarly to DQT, we assume $l_{s,d}<<2L_0$.
In EQT simulation, both transmons at source and destination nodes emit a microwave photon in each period % and each period takes into account duration time of microwave photon emission, entanglement generation through transduction and BSM \textcolor{blue}{(simulation time of 2 $ms$)}. 
and SeQUeNCe keeps track of the detectors' clicks to identify successful EPR pair distribution. As in DQT, the period is given by the reset time of the microwave cavity of the transducers (about $1 \mu s$).
Fig~\ref{fig:08} shows our simulation results where transducer conversion efficiency is set to three different values (0.8, 0.5, and 0.1). 
Similar to DQT, the number of trials $N$ for EQT simulations is fixed to 100. %is evaluated by the chosen $SE$, set to $5\%$.
The simulated values for different conversion efficiencies and the different proposed detectors are summarized in Tab.~\ref{tab:02}. 

\begin{table}[h]
\centering
\caption{EQT probabilities of detector clicks for different conversion efficiency values and detector types.}
\renewcommand{\arraystretch}{1.3}
\begin{tabular}{|c|c|c|c|}
\hline
\multicolumn{4}{|c|}{$\eta_\uparrow=0.8$} \\
\hline
\multirow{2}{*}{$p_c^{PNRD}$} & $\eta_d=1$ & 0.5 & 0.5 \\
\cline{2-4}
 & $\eta_d=0.25$ & 0.22 & 0.24 \\
\hline
\multirow{2}{*}{$p_c^{SPD}$} & $\eta_d=1$ & 0.75 & 0.75 \\
\cline{2-4}
 & $\eta_d=0.25$ & 0.23 & 0.28 \\
\hline
\multicolumn{4}{|c|}{$\eta_\uparrow=0.5$} \\
\hline
\multirow{2}{*}{$p_c^{PNRD}$} & $\eta_d=1$ & 0.32 & 0.34 \\
\cline{2-4}
 & $\eta_d=0.25$ & 0.32 & 0.32 \\
\hline
\multirow{2}{*}{$p_c^{SPD}$} & $\eta_d=1$ & 0.96 & 0.98 \\
\cline{2-4}
 & $\eta_d=0.25$ & 0.36 & 0.37 \\
\hline
\multicolumn{4}{|c|}{$\eta_\uparrow=0.1$} \\
\hline
\multirow{2}{*}{$p_c^{PNRD}$} & $\eta_d=1$ & 0.18 & 0.16 \\
\cline{2-4}
 & $\eta_d=0.25$ & 0.05 & 0.05 \\
\hline
\multirow{2}{*}{$p_c^{SPD}$} & $\eta_d=1$ & 0.19 & 0.17 \\
\cline{2-4}
 & $\eta_d=0.25$ & 0.05 & 0.06 \\
\hline
\end{tabular}
\label{tab:02}

\end{table}

%\begin{figure}[t!]
	%\centering
	%\begin{minipage}[c]{\linewidth}
	%	\centering
		%\includegraphics[width=1%\columnwidth]{Figures/Fig06.1.png}
		%\subcaption{}
		%\label{fig:06.1}
	%\end{minipage}
	%\hfill\hspace{3pt}
	%\begin{minipage}[c]{\linewidth}
		%\centering
		%\includegraphics[width=1\columnwidth]{Figures/Fig06.2.png}
		%\subcaption{}
		%\label{fig:06.2%}
	%\end{minipage}
	%\caption{Simulation results of EQTwith $\eta_{\uparrow}=0.5$.}
%	\label{fig:06}
%	\hrulefill
%\end{figure}

\begin{remark}
    The implicit assumption of the proposed theoretical analysis is that photons reaching the BSM node are indistinguishable \cite{HalBevGis-07, HubReiYon-17}. This assumption is taken into account in SeQUeNCe simulations with a simultaneous entanglement generation at both the source and the destination, the same length of the quantum channels (source-BSM and BSM-destination), and negligible attenuation of the optical channels. In an experimental setting, the synchronization that ensures the indistinguishability of photons is extremely challenging due to timing jitter and clock misalignment of different nodes~\cite{RamReiDan-25}. It will be crucial to analyze also the synchronization issue in future research.
\end{remark}

\section{Discussion}
\label{sec:05}

Despite the numerous technological advances, obtaining high conversion efficiency transducers is still a crucial challenge. 
In DQT strategy, a high value of conversion efficiency is strictly required in order to obtain a high value of probability of qubit transmission.
Furthermore, it is evident that low conversion efficiency values inevitably degrade the communication performances of this strategy as our simulation results confirm.
On the contrary, in the EQT strategy, the requirement on the conversion efficiency values is not so restrictive. Indeed it is sufficient to have $\eta_{\uparrow}>0$ for generating entangled pair at both source and destination.
Our simulation results show that transducers with an $\eta_{\uparrow}=0.1$ using the EQT strategy and real PNRD can distribute entanglement with 0.05 probability, while DQT under similar conditions is unable to transmit any photon.

It is worth to notice that in order to generate EPR pairs within the transducer with a beam splitter interaction, $\eta_{\uparrow}=0.5$ is required (as mentioned in a remark of Sec.\ref{sec:02.2}). Moreover, $\eta_{\uparrow}=0.5$ is also the condition for achieving the maximum  probability of EPR distribution. In contrast to what happens with DQT, a further incrementation of the conversion efficiency degraded the EQT performances.
For instance, when $\eta_{\uparrow}=0.8$ and using ideal PNRD, the probability of entanglement distribution decreases. 
Therefore, taking into account that obtaining high conversion efficiency is one of the main limitation factors for achieving quantum transduction, EQT relaxes this stringent limitation and constitutes a more viable strategy for quantum information transmission.
Moreover, our simulations results may suggest that it would be more productive to improve the efficiency of PNRD than improving the efficiency of quantum transducers above 0.5.

\section{Conclusion}
\label{sec:06}

In this paper, we presented two different communication models for quantum information transmission via quantum transduction: Direct Quantum Transduction (DQT) and Entanglement-based Quantum Transduction (EQT), and we evaluated the communication performances of the proposed strategies using the network simulator SeQUeNCe. 
%Our simulations constitute an experimental validation of theoretical results.
Under the same conditions ($\eta_\uparrow=0.5$) and assuming noiseless LOCC, the EQT strategy allows to have a probability of heralded distributed entanglement $p_e$ higher than the probability of successfully transferring the information qubit $p$ using DQT.
Indeed, the values obtained are consistent with the expected ones, giving us confidence in the accuracy of our model.

We implemented and open-sourced a quantum transducer module in SeQUeNCe along with additional hardware devices and protocols.
The software structure is modular and can be exploited for other strategies and larger network topologies. The transducer model can be easily extended by introducing additional noise within the transducer itself or in the proposed network, allowing higher accuracy of the model.
For these reasons, the implemented strategies in SeQUeNCe could be of significant interest to the scientific community.

\bibliographystyle{IEEEtran}
\bibliography{bibliography.bib}

\end{document}